\documentstyle[12pt,amssymbols]{article}

\def\d{{\rm d}}

\newtheorem{prop}{Proposition}[section]
\newtheorem{lem}{Lemma}[section]
\newtheorem{theo}{Theorem}[section]

\def\presentation{
\voffset -.5in
\hoffset -.19in
\oddsidemargin 0in
\evensidemargin 0in
\marginparwidth .75in
\marginparsep 7pt
\topmargin 0in
\headheight 12pt
\headsep .25in
\footheight 18pt
\footskip .35in
\textheight 9.5in
\textwidth 6.5in
\columnsep 10pt
\columnseprule 0pt
}

\presentation

\title {Construction of the classical $R$-matrices for the Toda
and Calogero models.}
 \author{J. Avan, O.Babelon, M.Talon.
 \thanks{L.P.T.H.E. Universit\'e Paris VI (CNRS UA 280),
 Box 126, Tour 16, $1^{er}$ \'etage,
 4 place Jussieu, F-75252 PARIS CEDEX 05} }
 \date{June 1993}

\begin{document}
\maketitle

\begin{abstract}
We use the definition
of the Calogero-Moser models as Hamiltonian reductions of geodesic motions on a
group
manifold to construct their $R$-matrices. In  the Toda case,
the analogous construction yields  constant $R$-matrices. By contrast, for
Calogero-Moser models they are dynamical objects.
\end{abstract}
\vfill
PAR LPTHE 93-31
\vfill
Work sponsored by CNRS, France.
\vfill
\newpage

\section{Introduction}

The unification by Faddeev and his school \cite{SkTaFa79,Fa82} of the classical
and
quantum inverse scattering
method with the Yang-Baxter equation was one of the
important achievements in the modern theory of integrable
systems. This lead to the concept of classical $R$-matrix \cite{Sk79,S1983}
which
encodes the Hamiltonian structure of Lax equations \cite{FT1986a}. \\
In the Lax representation of a dynamical system \cite{Lax 1968},
the equations of motion can be written
\begin{eqnarray}
\dot{L}=[M,L]
\nonumber
\end{eqnarray}
where $L$ and $M$ are elements of a Lie algebra ${\cal G}$.
The primary
interest of this representation is to provide us with conserved
quantities
\begin{eqnarray}
I_n = {\rm Tr} \;(L^n)
\nonumber
\end{eqnarray}
Integrability in the sense of Liouville \cite{A 1976,AM78} requires the
existence of
a sufficient number of conserved quantities in involution under
the Poisson bracket. Hence a dynamical system represented by a
Lax pair will be a natural candidate to integrability if the
$I_n$'s commute.\\
The commutation of the $I_n$'s is equivalent to the existence of
an $R$ matrix \cite{BaVia90}
\begin{eqnarray}
\{ L_1 , L_2 \} = [R_{12}, L_1] -[R_{21},L_2]
\label{matriced}
\end{eqnarray}
where the notation is as follows. If $e_i$ is a basis of the Lie
algebra ${\cal G}$, then $L=\sum_i L^i e_i$ and
\begin{eqnarray}
L_1=\sum_i L^i \;e_i \otimes 1 && L_2 = \sum_i L^i\; 1\otimes e_i \nonumber \\
\{ L_1 , L_2 \} &=& \sum_{i,j}\; \{L^i, L^j\}\; e_i \otimes e_j
\nonumber \\
R_{12}= \sum_{i,j} R^{ij}\;e_i \otimes e_j &&
R_{21} = \sum_{i,j} R^{ij}\;e_j \otimes e_i
\nonumber
\end{eqnarray}
{\bf Remarks.}\\
-- 1) The Lax operator $L$ and the $R$-matrix encode all the information
about the dynamical system. In particular when the Hamiltonian is chosen to be
$I_n$, the $M$ matrix of the corresponding flow reads
\begin{eqnarray}
M_n =-n\, {\rm Tr}_2\, R_{12}L_2^{n-1}
\label{RdH}
\end{eqnarray}
-- 2) The $R$-matrix in general is a non-constant function on the phase
space. The first examples of such $R$-matrices occurred in
\cite{Maillet 1985}.\\
-- 3) Since the above formula expresses only the involution
property of the eigenvalues of the Lax matrix $L$, every conjugate
matrix $L^g = g^{-1} L g$, where $g$ is any group valued function on
the phase space, also admits an $R$-matrix which can be explicitely
computed.\\
-- 4) Although this theorem guarantees the existence of an $R$-matrix if
one knows that the $I_n$'s are in involution, it does not provide a
practical way to find it. Experienced scholars in this domain
know that it is in general not an easy task to find an $R$-matrix,
and it is desirable to have a constructive way to obtain them.\\

We present here such a constructive scheme to obtain the
$R$-matrices for the Calogero-Moser models. The standard
Calogero-Moser model \cite{Cal75,Mos75,Cal76}
describes a set of $n$ particles submitted to the equations of
motion
\begin{eqnarray}
\ddot{q}_i = \sum_{j \neq i} {\cosh (q_i -q_j)
\over \sinh^3 (q_i -q_j)}
\nonumber
\end{eqnarray}
This model admits a Lax representation with
\begin{eqnarray}
L&=& p +\sum_{k \neq l} {i\over \sinh (q_k -q_l)} E_{kl}
\nonumber
\end{eqnarray}
where $p$ is a traceless diagonal matrix containing the
momenta. The dynamical $R$-matrix for this model was first found
in \cite{AvTa93}.\\

It is well known \cite{OlPe76,KaKoSt78,OlPe81} that both the Toda
models and the Calogero-Moser
models are obtained by Hamiltonian reduction of the geodesic motion
on the cotangent bundle $T^*G$ of a Lie group $G$. Denoting by $g,\;
\xi$ the coordinates on $T^*G$, we will see that the Poisson
structure on $T^*G$ implies the existence of an $R$-matrix for $\xi$
i.e.
\begin{eqnarray}
\{ \xi_1 , \xi_2 \} = [C_{12}, \xi_1] -[C_{21},\xi_2]
\nonumber
\end{eqnarray}
where $C_{12}$ is the quadratic Casimir element in ${\cal G}\otimes
{\cal G}$ and we have used an invariant bilinear form to identify
${\cal G}^*$ and ${\cal G}$. We shall then use the fact that in the
reduction process, the Lax matrix of the reduced system is expressed
in terms of $\xi$ by a formula of the type
\begin{eqnarray}
L= h \xi h^{-1}
\nonumber
\end{eqnarray}
where $h$ is some element in $G$. The remark then enables us to
compute the $R$-matrix of the reduced system.\\

In contrast with the Toda case,
where the $R$-matrix is a constant, this method gives us dynamical
$R$-matrices for the Calogero-Moser models. We recover the previously
known $R$-matrix for the standard Calogero-Moser model which
corresponds to the symmetric space $Sl(N,\Bbb C)/SU(n)$.\\

We will also consider a generalization  of the standard Calogero-Moser
model associated to the symmetric space $SU(n,n)/S(U(n) \times U(n))$.
The Hamiltonian of this system reads
\begin{eqnarray}
H={1\over 2}\sum_{i=1}^N p_i^2 +\sum_{i\neq j}^N\left\{ {1\over \sinh^2 (q_i
-q_j)}
+{1\over \sinh^2 (q_i +q_j)}\right\} +\sum_{i=1}^N {(1+\gamma)^2\over
\sinh^2(2q_i)}
\nonumber
\end{eqnarray}
where $\gamma$ is an arbitrary real coupling constant.
We obtain straightforwardly the $R$-matrix for this case
 which had eluded a direct computation
scheme.\\

The construction presented here applies to a Lax representation without
spectral
parameter. It is known that there exists another Lax representation for the
Calogero models depending on a spectral parameter \cite{Kri80}. The
corresponding
$R$-matrices were recently computed by E.K. Sklyanin \cite{Skl93}. It would be
interesting to find the equivalent scheme yielding these $R$-matrices.

\section{Hamiltonian reduction on cotangent bundles. }

\setcounter{equation}{0}

\subsection{Hamiltonian reduction.}

We begin by recalling some well-known facts concerning the Hamiltonian
reduction
of dynamical systems whose phase space is a cotangent bundle \cite{AM78}. Let
$M$ be a
manifold and $N=T^* M$ its cotangent bundle. $N$ is equipped with the canonical
1-form $\alpha$ whose value at the point $p \in T^* M$ is $\pi^* p$ where $\pi$
is
the projection of $N$ on $M$. Hence  any group of diffeomorphisms of $M$,
lifted naturally to $N$, namely $\phi \in {\rm Diff}(M)$ lifted to
$(\phi^*)^{-1} \in {\rm Diff}(N)$, leaves $\alpha$ invariant. We shall be
especially interested in the case in which a Lie group $G$ acts on $M$. Each
element $X \in \cal G$ (the Lie algebra of $G$) generates a vector field on $M$
that we shall denote $X.m$ at the point $m \in M$. It is the derivative of the
application $g \to g.m$ at the unit element of $G$. Since each map $m \to g.m$
is
a diffeomorphism of $M$ this lifts to a vector field on $N$ leaving $\alpha$
invariant. We shall also denote $X.p \in T_p(N)$ the value at $p \in N$ of this
vector field, so that the Lie derivative ${\cal L}_{X.p} \, \alpha$ of the
canonical
1-form vanishes.

Notice that $N$ is a symplectic manifold equipped with the
canonical 2-form
\begin{equation}
\omega = - {\rm d}\alpha. \label{omalp}
\end{equation}
To any function (or Hamiltonian) ${ \rm H}$
on $N$ we associate a vector field $X_{ \rm H}$ such that:
\begin{equation}
{\rm d} { \rm H} = i_{X_{ \rm H}}\, \omega \label{hx}
\end{equation}
and conversely since $\omega$ is non--degenerate. It is easy to find the
Hamiltonian associated to the above vector field $X.p,\; X \in \cal G$.
As a matter of fact we have $0={\cal L}_{X.p} \, \alpha={\rm d} i_{X.p}\alpha+
i_{X.p} {\rm d} \alpha$ hence:
\begin{equation}
{ \rm H}_X(p)=i_{X.p}\, \alpha = \alpha\,(X.p) \label{moment}
\end{equation}

For any two functions $F$, $G$ on $N$  one defines the Poisson bracket
$\{F,G\}$
as a function on $N$ by:
\begin{equation}
\{F,G\}=\omega\,(X_F,X_G)
\end{equation}
The Poisson bracket of the Hamiltonians associated to the group action has a
simple expression. This follows from the:
\begin{lem} \label{crochet}
For any $X$, $Y$  in $\cal G$ and $p\in N$ we have:
$$\omega\,(X.p,Y.p)=\alpha\,([X,Y].p) {\rm ~and~}
[X.p,Y.p]\equiv {\cal L}_{X.p}\,Y.p= -[X,Y].p$$
\end{lem}

\noindent{\it Proof.} By definition of the Lie bracket:
$${\cal L}_{X.p}\,Y.p={\d \over \d t} e^{-X.t}.\left(
Y.( e^{Xt} p) \right)_{\, | \, t=0}=-[X,Y].p$$
Then from the equations~(\ref{hx},\ref{moment}) and the invariance of $\alpha$
we get:
$$\omega\,(X.p,Y.p)=<\d {\rm H}_X , Y.p>={\cal L}_{Y.p}\,\alpha\,(X.p)
=\alpha\,([X,Y].p)$$
Noticing that $\omega\,(X.p,Y.p)=\{{ \rm H}_X,{ \rm H}_Y\}$
we have shown that the group action is Poissonnian, i.e.:
\begin{equation}
\{{ \rm H}_X,{ \rm H}_Y\}={ \rm H}_{[X,Y]} \label{2.hom}
\end{equation}

Obviously the application $X\in {\cal G} \to { \rm H}_X(p),\;
{\rm any}\; p \in N$, is
a linear map from ${\cal G}$ to the scalars and so defines an element
${\cal P}(p)$ of ${\cal G}^*$
which is called the momentum at $p \in N$. For any flow
induced by a Hamiltonian ${ \rm H}$ invariant under $G$ the momentum is
conserved (this is No\" ether's theorem):
$${\d \over \d t}<{\cal P}, X>\,=-\{{ \rm H},{ \rm H}_X\}=
{\cal L}_{X.p}\, { \rm H} =0,
\forall X \in \cal G$$
Hence in this case one can restrict oneself to the submanifold $N_\mu$ of $N$
with fixed momentum $\mu$ i.e. such that:
\begin{equation}
N_\mu ={\cal P}^{-1} (\mu) \label{2.nmu}
\end{equation}
assuming this is a well--defined manifold.

Due to  equation (\ref{moment})  and the invariance of $\alpha$
the action of the group $G$ on $N$ is
transformed by $\cal P$ into the coadjoint action of $G$ on ${\cal G}^*$
\begin{equation}
{\cal P}(g.p)(X) = \alpha\,(g.g^{-1}Xg.p) =
{\rm Ad}^*_{g} \,{\cal P}(p) (X)
\end{equation}
where the coadjoint action on an element $\xi$ of ${\cal G}^*$ is
defined as:
$$ {\rm Ad}^*_g \, \xi \, (X) = \xi \, (g^{-1} X g) $$
The stabilizer $G_\mu$ of $\mu \in {\cal G}^*$ acts on $N_\mu$. The reduced
phase space is precisely obtained by taking the quotient (assumed
well-behaved):
\begin{equation}
{\cal F}_\mu = N_\mu / G_\mu
\end{equation}
It is known that this is a symplectic manifold. As a matter of fact, for any
two
tangent vectors $\zeta,\, \eta$ at a point $f \in \cal F_\mu$ one defines:
$$\omega_f\,(\zeta,\eta)=\omega_p\,(\zeta',\eta')$$ where $\zeta',\eta'$ are
any
tangent vectors to $N_\mu$ projecting to $\zeta,\, \eta$, at some point $p \in
N_\mu$ above $f$. This definition is independent of the choices.

In the following we shall need to compute the Poisson bracket of functions
on ${\cal F}_\mu$. These functions are conveniently described as $G_\mu$
invariant functions on $N_\mu$. To compute their Poisson
bracket we first extend them arbitrarily in the vicinity of $N_\mu$.
Two  extensions differ by a function vanishing on $N_\mu$. The difference
of the Hamiltonian vector fields of two such extensions
is controlled by the following:
\begin{lem}
Let $f$ be a function defined in a vicinity of $N_\mu$ and vanishing on
$N_\mu$.
Then the Hamiltonian vector field $X_f$ associated to $f$ is tangent to the
orbit $G.p$ at any point $p \in N_\mu$.
\end{lem}

\noindent{\it Proof.} The subvariety $N_\mu$ is defined by the equations
${\rm H}_{X_i}=\mu_i$ for some basis $X_i$ of $\cal G$.
Since $f$ vanishes on $N_\mu$ one can write $f=\sum\, ({\rm H}_{X_i}-\mu_i)f_i$
for some functions $f_i$ defined in the vicinity of $N_\mu$. For
any tangent vector $v$ at a point $p\in N_\mu$ one has:
$$<\d f(p),v>=\sum_i <\d {\rm H}_{X_i}(p),v>f_i(p)=
\omega\, (\,\sum_i\, f_i(p)\,X_i.p,v\,)$$
since the  Hamiltonian vector field associated to ${\rm H}_{X_i}$ is $X_i .p$
and
$\sum ({\rm H}_{X_i}-\mu_i)\, \d f_i$ vanishes on $N_\mu$. Hence
$X_f=\sum \, f_i(p)\,X_i.p \in {\cal G}.p$.

As a consequence of this lemma we have a method to compute the reduced Poisson
bracket. We take two  functions defined on $N_\mu$ and invariant under $G_\mu$
and extend them arbitrarily. Then we compute their Hamiltonian vector fields on
$N$
and project them on the tangent space to $N_\mu$ by adding a vector tangent
to the orbit $G.p$. These projections are independent of the extensions and the
reduced Poisson bracket is given by the value of the symplectic form on $N$
acting on them.
\begin{prop}
At each point $p \in N_\mu$ one can choose a vector $V_f.p \in {\cal G}.p$ such
that $X_f + V_f.p \in T_p(N_\mu)$ and $V_f.p$ is determined up to a vector
in ${\cal G}_\mu .p$.
\end{prop}

\noindent{\it Proof.} Let us notice that the symplectic orthogonal of ${\cal
G}.p$
is exactly $T_p(N_\mu)$. This is because  $\omega(\xi, X.p)=0$ for any
$X \in \cal G$
means $\d {\rm H}_X(\xi)=0$ hence $\xi \in T_p(N_\mu)$ since $N_\mu$ is
defined by the equations ${\rm H}_X=\mu(X)$.
Hence $T_p(N_\mu)\cap {\cal G}.p={\cal G}_\mu.p$  is the kernel of
the symplectic form restricted to ${\cal G}.p$.
We want to solve $\omega\,(X_f+V_f.p,X.p)=0,\,\forall X \in \cal G$ i.e.
\begin{equation}
\chi\,(X,V_f)=(X.p).f,\; \forall X \in \cal G
\label{eqvv}
\end{equation}
where we have introduced:
\begin{equation}
\chi\,(X,Y)= \omega\,(X.p,Y.p) \label{chip}
\end{equation}
Let us remark that this form on $N_\mu$ only depends on the momentum $\mu$
since assuming that the the group action is Poissonian~(\ref{2.hom}) we have:
\begin{equation}
\chi\,(X,Y)=\{{\rm H}_X,{\rm H}_Y\}={\rm H}_{[X,Y]}=
{\cal P}\,([X,Y])=\mu\,([X,Y])
\label{chimu} \end{equation}

Since ${\cal G}_\mu=\{X\in {\cal G}\,|\,{\rm ad}^*_X \,\mu=0\}$ $\chi$
defines a non--degenerate skew--symmetric bilinear form on
${\cal G}/{\cal G}_\mu$.
Finally $\chi$ induces a canonical isomorphism $\hat{\mu}\colon
{\cal G}/{\cal G}_\mu \to ({\cal G}/{\cal G}_\mu)^*$ by setting
$(\hat{\mu}(Y))(X) = \chi\,(X,Y)$.
Since $f$ is ${\cal G}_\mu$--invariant on $N_\mu$
the right--hand side of equation~(\ref{eqvv}) defines an element $\lambda_f:\;
X\to (X.p)\cdot f$ in $({\cal G}/
{\cal G}_\mu)^*$. Then $V_f$ may be seen as an element of ${\cal G}/{\cal
G}_\mu$
given by $\hat{\mu}^{-1} (\lambda_f)$.

For any such functions $f,g$ the reduced Poisson bracket is given by:
\begin{equation}
\{ f,g \}_{\rm reduced}=\omega\,(X_f+V_f.p,X_g+V_g.p)=
\{ f,g \}-\omega\,(V_f.p,V_g.p) \label{poisred}
\end{equation}
Notice that if $f_{\, |\, N_\mu}=0$ we have $X_f+V_f.p\in  {\cal G}_\mu.p$
hence $\{ f,g \}_{\rm reduced}=
\omega\,(X_f+V_f.p,X_g+V_g.p)=0$ so that equation~(\ref{poisred}) indeed
defines a Poisson bracket on the reduced phase space. From eq.~(\ref{chimu})
we have:
$$\omega\,(V_f.p,V_g.p)= \,<\mu,[V_f,V_g]>$$
which can be further simplified by substituting $X=V_g$ in
equation~(\ref{eqvv}).
By antisymmetrization one gets:  $\omega\,(V_f.p,V_g.p)=
(1/2)\,(\,(V_f.p)\cdot g-(V_g.p)\cdot f\,)$.

Assuming in particular that $f$ and $g$ are $G$--invariant extensions of our
given
functions on $N_\mu$ it is obvious that $\{f,g\}$ is $G$--invariant
(invariance of $\omega$) hence its restriction to $N_\mu$ is
$G_\mu$--invariant and independent of the choices. Moreover
the associated Hamiltonian vector fields $X_f,\, X_g$ are tangent
to $N_\mu$ since the $G$--invariance of $f$ implies:
$$0=\d f\,(X.p)=\omega\,(X_f,X.p)=-\d {\rm H}_X(X_f),\;X\in\cal G$$
therefore the functions ${\rm H}_X$ are constant along $X_f$ i.e.
$X_f$ is tangent to $N_\mu$. As a result
the symplectic form defined above on ${\cal F}_\mu$ yields the
same Poisson brackets for $f$ and $g$ as computed by this method.

We have shown the:
\begin{prop}
The reduced Poisson bracket of two functions on ${\cal F}_\mu$ can be
computed using any extensions $f,\, g$ in the vicinity of $N_\mu$
according to:
\begin{equation}
\{f,g\}_{\rm reduced}=\{f,g\}+{1\over 2}\left(
(V_g.p).f -(V_f.p).g
\right) \label{bienreduit}
\end{equation}
This is equivalent to  the Dirac bracket.
\end{prop}

\subsection{The case $N=T^*G$}

If $M=G$ is a Lie group, one can use the left translations to
identify $N=T^*G$ with $G\times {\cal G}^*$.
\begin{equation}
\omega \in T^*_g(G) \longrightarrow
(g,\xi)\mbox{~~~where~~}\omega=L^*_{g^{-1}}\xi
\end{equation}
If $(v,\kappa)$ is a  vector tangent to $T^*G$ at the point $(g,\xi)$, the
canonical
1-form, invariant under both left and right translations, is given by:
\begin{equation}
\alpha \, (v,\kappa)= \xi \, (g^{-1}\cdot v ) \label{can1form}
\end{equation}
The right action of $G$ on $G$ produces left invariant vector fields $g.X, \;
X\in \cal G$ which can be lifted to $T^*G$.
The associated Hamiltonians are simply:
\begin{equation}
{ \rm H}_X= \xi (X)
\end{equation}
The equation (\ref{2.hom}) for right actions becomes
$\{{ \rm H}_X,{ \rm H}_Y\}= -{ \rm H}_{[X,Y]}$ as may be seen by considering
the left action $g' . g = gg'^{-1}$ hence:
\begin{equation}
\{ \xi (X), \xi (Y) \} = - \xi ([X,Y]), \label{poi1}
\end{equation}
i.e. the Poisson bracket of the $\xi$'s is just the Kirillov bracket.
Moreover, since $H_X$ generates a  right translation, we have
$\{ {\rm H}_X , g \}=\omega\,(g.X,X_g)=-\d g(g.X)=-g.X$ hence:
\begin{equation}
\{ \xi (X) ,g \} = - g\;X \label{poi2}
\end{equation}
Finally, we have a complete description of Poisson brackets with:
\begin{equation}
\{ g , g \} =0 \label{poi3}
\end{equation}

Geodesics on the group $G$ correspond to  left translations of
1-parameter groups (the
tangent vector is transported parallel to itself), therefore
\begin{equation}
{d\over dt}(g^{-1}\dot{g}) =0
\end{equation}
This is a Hamiltonian system whose Hamiltonian is:
\begin{equation}
{ \rm H} = {1 \over 2}\; (\xi,\xi)
\end{equation}
where we have identified ${\cal G}^*$ and $\cal G$ through the invariant
Killing metric.

Notice that ${\rm H}$ is bi--invariant, so one can attempt to
reduce this dynamical system using  Lie subgroups $H_L$
and $H_R$ of $G$ of Lie algebras ${\cal H}_L$
and ${\cal H}_R$, acting respectively on the left and on the right on $T^*G$
in order to obtain a non--trivial result.

Using the coordinates $(g,\xi)$ on $T^*G$  this action reads:
$$(\;(h_L,h_R),(g,\xi)\;)\to (h_L g h_R^{-1},{\rm Ad}^*_{h_R} \xi)$$
We have written this action as a left action on $T^*G$, in order to
apply the formalism developed in Section~(2.1). The infinitesimal
version of this action is given by:
\begin{equation}
(X_L,X_R)\cdot (g,\xi)=(X_L .g -g.X_R,[X_R,\xi])
\label{action}
\end{equation}
so that the corresponding Hamiltonian can be written:
$${\rm H}_{(X_L,X_R)}(p)=\alpha\,((X_L,X_R).p)=\,
<g \xi g^{-1},X_L>-<\xi,X_R>$$
This means that the moments are:
\begin{equation}
{\cal P}^L(g,\xi ) = P_{{\cal H}^*}\, {\rm Ad}^*_{g}\, \xi \quad
{\cal P}^R(g,\xi ) = - P_{{\cal H}^*}\, \xi \label{1.mom} \quad
{\cal P}=({\cal P}^L,{\cal P}^R)
\end{equation}
where we have introduced the projector on ${\cal H}^*$ of forms in
${\cal G}^*$ induced by the restriction of these forms to $\cal H$.

\section{The Toda model}

\setcounter{equation}{0}

Let us now apply this construction to obtain the $R$--matrix of the Toda
models.
This was first done by Ferreira and Olive~\cite{FeOl85}, but we wish to present
here
a short discussion of this case since it is a very simple illustration of the
general scheme which we shall use in the more complicated case of Calogero
models.

\subsection{Iwasawa decomposition.}

Let $G$ be a complex simple Lie group with Lie algebra ${\cal G}$. Let $\{
H_i  \}$ be the generators of a Cartan subalgebra $\cal H$,
and let $\{ E_{\pm \alpha} \}$ be the corresponding root vectors,
chosen to form a Weyl
basis, i.e. all the structure constants are real. The real normal form of
$\cal G$ is the real Lie algebra ${\cal G}_0$ generated over $\Bbb{R}$ by
the $H_i$ and $E_{\pm\alpha}$.  Let $\sigma$ be the Cartan involution:
$\sigma (H_i)=-H_i$, $\sigma (E_{\pm \alpha}) = -E_{\mp \alpha}$. The
fixed points of $\sigma $ form a Lie subalgebra $\cal K$
of ${\cal G}_0$ generated by $\{ E_\alpha - E_{-\alpha} \}$.
We have the  decomposition:
$${\cal G}_0={\cal K} \oplus {\cal M}$$
where $\cal M$ is the real vector space generated by the
$\{ E_\alpha + E_{-\alpha} \}$
and the $\{H_i  \}$. Notice that due to the choice of the real normal form
${\cal A}={\cal H}\cap{\cal G}_0$ is a maximal abelian subalgebra of
${\cal G}_0$ and it is entirely contained in $\cal M$. Finally we need the
real nilpotent subalgebras ${\cal N}_\pm$ generated respectively by the
$\{ E_{\pm \alpha} \}$.

Let $G_0$ be the connected Lie group corresponding to the Lie algebra
${\cal G}_0$, and similarly $K$ corresponding to $\cal K$ and
$N_\pm$ corresponding to ${\cal N}_\pm$. Notice that $G_0/K$ is
a symmetric space of the non--compact type. Finally the Cartan
algebra $\cal A$ exponentiates to $A$.

The connected Lie group $G_0$ admits the following Iwasawa
decomposition: $$G_0 \simeq N_+ \times A \times K {\rm
{}~as~a~manifold~}$$ that is any element $g$ in $G_0$ can be written
uniquely $g=nQk$.  We shall perform the reduction of the geodesic
motion on $T^*G_0$ by the action of the group $N_+$ on the left and $K$
on the right.

\subsection{The moment map.}

The reduction is obtained by a suitable choice of the momentum. We take:
\begin{eqnarray}
 P_{{\cal K}^*}(\xi)&=&\mu^R =0 \label{momenttoda1} \\
 P_{{\cal N}_+^*}(g^{-1}\xi g) &=& \mu^L = \sum_{\alpha ~{\rm simple}}
E_{-\alpha}
\label{momenttoda2}
\end{eqnarray}
where we have identified ${\cal N}_+^*$ with ${\cal N}_-$ through the
Killing form.

The isotropy group $G_\mu$ is $N_+ \times K$. This is obvious for the
right component since $\mu^R=0$. The isotropy group of $\mu^L$ is
by definition the set of elements $g \in N_+$ such that:
$$ < \mu^L  ,g^{-1}Xg > \,= \,<\mu^L ,X> ~\forall X \in {\cal N}_+ $$
Since $\mu^L$ only contains roots of height -1, the only
contribution to $<\mu^L,X>$ comes from $X^{(1)}$, the level one
component of X. But $(g^{-1}Xg)^{(1)} =X^{(1)} ~\forall g \in N_+$.
Hence the isotropy group of $\mu^L$ is $N_+$ itself.

\subsection{The submanifold $N_\mu$.}

Let us first compute the dimension of the reduced phase space. Let
$d={\rm dim }\,{\cal G}_0$ and $r={\rm dim }\,\cal A$. Then we have:
$${\rm dim }\,{\cal K}={\rm dim }\,{\cal N_+}={d-r \over 2}\quad
{\rm dim }\,T^*G_0=2d$$
The dimension of the submanifold $N_\mu$ defined by the
equations~(\ref{momenttoda1},
\ref{momenttoda2})
is ${\rm dim }\,N_\mu=2d-{\rm dim }\,{\cal K}-{\rm dim }\,{\cal N_+}=d+r$
and the dimension of the reduced phase space is
${\rm dim }\,{\cal F}_\mu={\rm dim }\,N_\mu-{\rm dim }\,
{\cal K}-{\rm dim }\,{\cal N_+}=2r$
which is the correct dimension of the phase space of the Toda chain.

We now construct a section of the bundle $N_\mu$ over ${\cal F}_\mu$. Since
the isotropy group is the whole of $N_+ \times K$ any point $(g,\xi)$ of
$N_\mu$
can be brought to the form $(Q,L)$ with $Q \in A$ (due to the Iwasawa
decomposition)
by the action of the isotropy group. In this subsection we shall identify
${\cal G}_0$ and ${\cal G}_0^*$ under the Killing form for convenience.
Equation~(\ref{momenttoda1}) implies that $L \in \cal M$ which is the
orthogonal
of $\cal K$. Thus we can write:
$$L=p+\sum_\alpha \, l_\alpha (E_\alpha+E_{-\alpha}),\quad p\in \cal A$$
Inserting this form into equation~(\ref{momenttoda2}) and setting
$Q=\exp(q)$ we get:
$$P_{N_-}\,Q^{-1}LQ = \sum_\alpha \,l_\alpha \exp \alpha(q) E_{-\alpha}=
\sum_{\alpha ~{\rm simple} }\, E_{-\alpha}$$
hence $l_\alpha=\exp(-\alpha(q))$ for $\alpha$ simple and $l_\alpha=0$
otherwise.
We have obtained the standard Lax matrix of the Toda chain:
\begin{equation}
L=p\,+\sum_{\alpha ~{\rm simple} }\, e^{-\alpha(q)}\,(E_{\alpha}+E_{-\alpha})
\label{laxtoda}
\end{equation}
The set of the $(Q,L)$ is obviously a $(2r)$--dimensional subvariety $\cal S$
of $N_\mu$ forming a section of the above--mentioned bundle. This means that
for
any point $(g,\xi)$ in $N_\mu$ one can write uniquely $g=nQk$ and
$\xi=k^{-1}Lk$.

\subsection{The $R$--matrix of the Toda model.}

The function $L(X)$ (for any $X \in\cal M$)
defined on the section $\cal S$ has a uniquely defined
extension on $T^*G_0$, invariant under the action of the group $N_+ \times K$:
\begin{equation}
F_X(g,\xi)=\,<\xi,k^{-1}Xk > \label{fxtoda}
\end{equation}
where $k=k(g)$ is uniquely determined by the Iwasawa decomposition $g=nQk$.
In this situation equation~(\ref{bienreduit}) has no term
$\omega\,(V_f.p,V_g.p)$
corresponding to a projection on $TN_\mu$ and we have simply:
$$\{ L(X), L(Y)\}=\{F_X,F_Y\}$$
This is evaluated immediately with the help of equations~(\ref{poi1},
\ref{poi2}, \ref{poi3})
and leads on the section $\cal S$ to:
$$\{ L(X), L(Y)\}=-L([X,Y])+<L,[X,{\nabla_g k }(Y)\vert_{\cal S}]
+[{\nabla_g k}(X)\vert_{\cal S},Y]>$$
where the derivatives of $k$  are defined as:
$$\nabla_g k(X)={\d \over \d t}k(g \exp(tX))_{|t=0}$$
Notice that $[X,Y]\in \cal K$ since $G_0/K$ is a symmetric space hence
$L([X,Y])=0$.
{}From this equation follows immediately an $R$--matrix structure for the Toda
system
given by:
$$RX={\nabla_g k }(X)\vert_{\cal S}$$

We can compute the derivatives as follows: due to the Iwasawa
decomposition we have uniquely
$$QX=X_+ Q + X_a Q + Q X_K,\quad X_+\in {\cal N}_+,\;X_a \in {\cal A},\;
X_K \in \cal K$$
and ${\nabla k_g }(X)\vert_{\cal S}=X_K$.
Then multiplying this equation by $Q^{-1}$ on the left and noticing that
$Q^{-1}X_+ Q \in {\cal N}_+$ we get $X_K=\sum_\alpha
x_\alpha(E_\alpha-E_{-\alpha})$
when $X=\sum_\alpha x_\alpha(E_\alpha+E_{-\alpha})+\sum_i x_i H_i$.

In the dualized formalism defined by $RX={\rm Tr}_2 (R_{12}. {\bf 1}\otimes X)$
this $R$--matrix reads:
\begin{eqnarray}
R_{12}= {1 \over 2}
\sum_{\alpha > 0}\,( E_\alpha-E_{-\alpha})\otimes ( E_\alpha+E_{-\alpha})
\label{lanotre}
\end{eqnarray}

The classical form of the $R$-matrix for the Toda model reads
\cite{BD82,OT83,BD84}
$$R^{\rm \; standard}_{12}=\sum_{\alpha > 0}\,E_\alpha \otimes E_{-\alpha} -
E_{-\alpha}\otimes E_\alpha $$
Since $L$ in (\ref{laxtoda}) is by construction invariant under the
Cartan automorphism, one can write the Poisson brackets of $L\otimes
{\bf 1}$ with ${\bf 1}\otimes L$ using the $R$-matrix $R_{12}^{\;\sigma}=
\sigma \otimes {\bf 1}\; R^{\rm\; standard}_{12}$. Adding the two
$R$-operators gives back (\ref{lanotre}). This construction is an
application of a general formalism used for instance in the
construction of rational multipoles and trigonometric $R$-matrices
\cite{BD84,ReF83,AvTa91}.

\section{The Calogero models}

\setcounter{equation}{0}

Olshanetski and Perelomov~\cite{OlPe76} have shown that the Calogero--Moser
models can
be obtained by applying  a Hamiltonian reduction to the geodesic motion on
some suitable symmetric space.

\subsection{Symmetric spaces.}

Let us consider an involutive automorphism $\sigma$ of a simple Lie group $G$
and
the subgroup ${ H}$ of its fixed points. Then $H$ acts on the right on $G$
defining a
principal fiber bundle of total space $G$ and base $G/H$, which is a global
symmetric space. Moreover $G$ acts on the left on $G/H$ and in particular so
does
$H$ itself. We shall consider the situation described in Section~(2.2) when
$H_L=H_R=H$. The Hamiltonian of the geodesic flow on $G/H$ is
invariant under the $H$ action allowing to construct the Hamiltonian reduction
which under suitable choices of the momentum leads to the Calogero--Moser
models.
As a matter of fact since the phase space of the Calogero model is non compact
one has to start from a non compact Lie group $G$ and quotient it by a
maximal compact subgroup $H$ so that the symmetric space $G/H$ is of the
non compact type.

The derivative of $\sigma$ at the unit element of $G$ is an involutive
automorphism
of $\cal G$ also denoted $\sigma$. Let us consider its eigenspaces $\cal H$ and
$\cal K$  associated with the eigenvalues $+1$ and $-1$ respectively. Thus we
have a decomposition:
\begin{equation}
{\cal G}={\cal H} \oplus {\cal K} \label{ghk}
\end{equation}
in which $\cal H$ is the Lie algebra of $H$ which acts by inner automorphisms
on
the vector space $\cal K$ ($h{\cal K}h^{-1}=\cal K$).

 Let $\cal A$ be a
maximal commuting set of elements of $\cal K$. It is called a Cartan
algebra of the symmetric space $G/H$. It is known that every element in
$\cal K$ is conjugated to an element in $\cal A$ by an element of $H$.
Moreover $\cal A$ can be extended to a maximal commutative subalgebra of
$\cal G$ by adding to it a suitably chosen
abelian subalgebra $\cal B$ of $\cal H$. We shall
use the radicial decomposition of $\cal G$ under the abelian
algebra
${\cal A}$:
\begin{equation}
{\cal G}={\cal A}\bigoplus{\cal B}\bigoplus_{e_\alpha,\; \alpha \in
\Phi}\,\Bbb{R} e_\alpha
\label{radix}
\end{equation}

As a matter of fact the set of $\alpha \in \Phi$,
is a non reduced root system in ${\cal A}^*$ known as the root
system of the symmetric space $G/H$. Hence the root spaces are not
generically of dimension one. In the following $\sum_\alpha$
will denote the $\sum_{e_\alpha,\; \alpha \in \Phi}$. Equivalently we have the:
\begin{prop}
If $e_\alpha \neq 0$ is a root vector associated to the root $\alpha$,
$\sigma(e_\alpha)$ is a root vector associated to some other root denoted
$\sigma(\alpha)$. Then $\Phi$ is the disjoint union of a subset $\Phi'$
and $\sigma(\Phi')$.
\end{prop}

\noindent{\it Proof.} The root vector $e_\alpha$ is defined by the equation
$[q,e_\alpha]=\alpha(q) e_\alpha$, $ q\in{\cal A}$.
Applying to it the automorphism $\sigma$ one gets:
\begin{equation}
[q,\sigma(e_\alpha)]=\sigma(\alpha) \sigma(e_\alpha),\;
\sigma(\alpha)(q)=\alpha(\sigma(q)) \label{commut}
\end{equation}
This shows that $\sigma(e_\alpha)$ is a non zero root vector associated
to the linear form $\sigma(\alpha)$ which therefore belongs to $\Phi$.
Obviously $\sigma$ acts as an involutive bijection of $\Phi$ allowing
to separate it into $\Phi'$ and $\sigma(\Phi')$.

These decompositions of $\cal G$ exponentiate to similar decompositions of $G$.
First $G=KH$ where $K=\exp ({\cal K})$. It is known that
for a simply connected Lie group $G$ and
a non--compact symmetric space $G/H$, $K$ is diffeomorphic to $G/H$
and $K \times H \to G$ is a diffeomorphism (uniqueness of the
so--called Cartan decomposition).

Then $A=\exp ({\cal A})$ is a maximal
totally geodesic flat submanifold of $G/H$ and any element of $K$ can be
written as $k=h Q h^{-1}$ with $Q \in A$ and $h \in H$. It  follows that any
element of $G$ can be written as $g=h_1 Q h_2$ with $h_1,h_2 \in H$.

Of course this decomposition is  non unique. This non--uniqueness
is described  in the following:
\begin{prop} \label{unicite}
If $g=h_1Qh_2=h'_1Q'h'_2$ we have: $h'_1=h_1d^{-1}h_0^{-1}$,
$h'_2=h_0dh_2$ and $Q'=h_0Qh_0^{-1}$ where $d\in \exp({\cal B})=B$
and $h_0\in H$ is a representative of an element of the Weyl group
of the symmetric space. So if we fix $Q=\exp(q)$ such that $q$ be in a
fundamental Weyl chamber, the only ambiguity resides in the element $d \in B$.
\end{prop}

\noindent{\it Proof.} Settinq $h={h'}_1^{-1}h_1$ and
$h'=h'_2h_2^{-1}$ the equation reads: ${h'}^{-1}hQ={h'}^{-1}Q'h'\in
K$. By uniqueness of the Cartan decomposition ${h'}^{-1}h=1$ hence
$Q'=hQh^{-1}$. The adjoint action of h sending the centralizer of $Q$
to that of $Q'$ (when $Q$ hence $Q'$ are assumed regular), both being
equal to the maximal ``torus'' $AB$, we see that $h$ is in the
normalizer of $AB$.  But it is known that the quotient of this
normalizer in $H$ by the centralizer $AB$ is the so--called Weyl
group of the symmetric space $G/H$. Hence we can write $h=h_0d$ where
$h_0\in H$ is a representative of this quotient, and
$d \in B$. The conclusion follows.

\subsection{The moment map}

The reduction is obtained by an adequate choice of the momentum
$\mu=(\mu^L,\mu^R)$ such that ${\cal P}=\mu$. We take $\mu^R=0$
so that the isotropy group of the right component is $H_R$ itself.

The choice of the moment $\mu^L$ is of course of crucial
importance. We will consider $\mu^L$'s such that
\begin{itemize}
\item  their isotropy group $H_\mu$ is a maximal proper Lie subgroup of $H$,
so that the phase space of the reduced system be of minimal dimension but
non trivial.
\item In order to ensure the unicity of the decomposition introduced
in the Proposition~(\ref{unicite})
on $N_\mu$ we shall need:
\begin{equation}
{\cal H}_\mu \cap {\cal B}=\{ 0\}  \label{2.cap}
\end{equation}
Obviously ${\cal B}$ is an isotropic subspace of the skew--symmetric form
$\chi$ introduced in~(\ref{chip}). We shall require that it is a maximal
isotropic subspace. We choose a complementary maximal isotropic subspace
$\cal C$ so that
\begin{equation}
{\cal H}={\cal H}_\mu \oplus {\cal B} \oplus \cal C \label{hbc}
\end{equation}
and $\chi$ is a non--degenerate skew--symmetric bilinear form on
${\cal B} \oplus \cal C$, hence ${\rm dim }\,{\cal B}={\rm dim }\,{\cal C}$.
Notice that $\cal C$ is defined up to a symplectic transformation
preserving $\cal B$.
\item The reduced phase space ${\cal F}_\mu$ has dimension $2\,{\rm dim}\,
{\cal A}$
\end{itemize}
This is a constraint on the choice of $\mu$
that will be verified in the specific examples below.

\subsection{The submanifold $N_\mu$}

We now give an explicit description of $N_\mu$ i.e. we construct a section
$\cal S$ of the bundle $N_\mu$ over ${\cal F}_\mu$ so that one can write:
\begin{equation}
N_\mu=H_\mu {\cal S}H \label{ndec}
\end{equation}
To construct this section we  take a point $Q$ in $A$ and
an $L \in {\cal G}^*$ such that the point $(Q,L)$ is in $N_\mu$.
In this subsection we shall for convenience identify $\cal G$ and
${\cal G}^*$ under the Killing form assuming that $G$ is semi--simple.
Moreover since the automorphism $\sigma$ preserves the Killing form,
$\cal H$ and $\cal K$ are orthogonal, and $P_{{\cal H}^*}$ reduces to
the orthogonal projection on $\cal H$.
Since $\mu^R=0$ we have $L\in {\cal K}$ and one can write:
\begin{equation}
L=p+\sum_{e_\alpha,\;\alpha \in \Phi'} l_\alpha\, (e_\alpha - \sigma(e_\alpha))
\label{2.L}
\end{equation}
where $p \in \cal A$.
{}From equation (\ref{1.mom}) one gets:
$$\mu^L= P_{{\cal H}} \left(
p + \sum_\alpha l_\alpha\, (Q e_\alpha Q^{-1} - Q \sigma(e_\alpha) Q^{-1})
\right)$$
Since $Q = \exp(q),\; q\in \cal A$ we have  $Qe_\alpha Q^{-1}=\exp(\alpha(q))
 e_\alpha$ and similarly $Q\sigma(e_\alpha) Q^{-1}=\exp(-\alpha(q))
\sigma(e_\alpha)$ by exponentiating equation~(\ref{commut}).

Then the above equation becomes:
\begin{equation}
\mu^L=\sum_{\alpha} l_\alpha \sinh \alpha(q)\, (e_\alpha +\sigma(e_\alpha))
\label{2.mu}
\end{equation}
One can choose the momentum of the form:
$\mu^L = \sum_\alpha g_\alpha (e_\alpha + \sigma(e_\alpha))$ namely $\mu^L$
has no component in $\cal B$, where the $g_\alpha$ are such
that $H_\mu$ is of maximal dimension (we shall see that it essentially
fixes them, and obviously if $g_\alpha \neq 0$ for any $\alpha$
equation~(\ref{2.cap}) is automatically satisfied) and we have shown the:
\begin{prop} \label{laxpair}
The couples $(Q,L)$ with $Q =\exp(q)$ and
$$L=p+\sum_{\alpha }\,{g_\alpha \over \sinh \alpha(q)}\,
(e_\alpha-\sigma(e_\alpha))$$
with $p,q \in \cal A$ form a submanifold in $N_\mu$ of dimension $2\,{\rm
dim}\,
{\cal A}$.
\end{prop}

Notice that $L$ is just the Lax operator of the Calogero model and that
the section $\cal S$ depends of $2\,{\rm dim}\,{\cal A}$ parameters in an
immersive way.
Hence one can identify $N_\mu$ with the set of orbits of $\cal S$
under $H_\mu \times H$ i.e. the set of points
$(g=h_1Qh_2,\xi=h_2^{-1}Lh_2)$ with $h_1\in H_\mu$ and $h_2\in H$
{\em uniquely defined} due to condition~(\ref{2.cap}). The variables
$p$ and $q$ appearing in $Q$ and $L$ are the dynamical variables of
the Calogero model.

Let us remark that the reduced symplectic structure on ${\cal F}_\mu$ may
be seen as the restriction on the section $\cal S$ of the symplectic form
$\omega$ on $N$. According to equation~(\ref{can1form}) the restriction
to $\cal S$ of the canonical 1--form is $<L,\d q>={\rm Tr}\,(p \d q)$
since the root vectors $e_\alpha$ are orthogonal to $\cal A$ under the
Killing form. Hence the coordinates $(p,q)$ form a pair of canonically
conjugate variables.

\subsection{The $R$--matrix of the Calogero model}

We want to compute the Poisson bracket of the functions on ${\cal F}_\mu$
whose expressions on the section $\cal S$ are $L(X)$ and $L(Y)$ for
$X,Y \in \cal K$. These functions have uniquely defined $H_\mu \times H$
invariant extensions to $N_\mu$ given respectively by:
$$F_X(g,\xi)=<\xi,h_2^{-1}Xh_2>,\;F_Y(g,\xi)=<\xi,h_2^{-1}Yh_2>
{\rm ~where~} g=h_1Qh_2$$
Notice that $h_2$ is a well--defined function of $g$ in $N_\mu$ due to
condition~(\ref{2.cap}). According to the prescription given in the
section~(2.1) we choose extensions of these functions in the vicinity
of $N_\mu$. We {\em define} these extensions at the point $p=(g,\xi) \in T^*G$
by the {\em same} formulae in which $h_2$ is chosen to be a function
depending {\em only} on $g$ and reducing to the above--defined $h_2$
when $p \in N_\mu$.
Because of the non--uniqueness of the decomposition $g=h_1Qh_2$
outside of $N_\mu$ one cannot assert that the functions $F_X,\,F_Y$ are
invariant under the action of $H\times H$ and we must appeal to the
general procedure to compute the reduced Poisson brackets.

According to the theory
developed in section~(2.1) it is necessary to compute the projection on
$T(N_\mu)$ of the Hamiltonian vector field associated to a function
$F_Z(g,\xi)=\xi(h_2^{-1}Zh_2)$. In order to compute the corresponding
vector $V_F$ at the point $p=(g,\xi)$ it is convenient to consider only left
action of the group $H\times H$ in the forms given in equation~(\ref{action}).
Then equation~(\ref{eqvv}) becomes:
\begin{eqnarray}
<\xi,g^{-1}[X_L,V_L]g-[X_R,V_R]>&=&<\nabla_g F_Z,g^{-1}X_Lg-X_R>\nonumber\\
&&+<[X_R,\xi],\nabla_\xi F_Z> \label{generale}
\end{eqnarray}
where the $F$-derivatives are defined as:
$$<\nabla_g F,X>={\d \over \d t}F(g \exp(tX),\xi)_{|t=0}\quad
<\eta,\nabla_\xi F>={\d \over \d t}F(g,\xi+t\eta)_{|t=0}$$

For the above--defined function $F_Z$ these derivatives are immediately
calculated and the equation~(\ref{generale}) becomes:
\begin{eqnarray}
\lefteqn{<\xi,g^{-1}[X_L,V_L]g-[X_R,V_R]>\,=}
\nonumber\\
&&<\xi,h_2^{-1}\,[Z,\nabla_g
h_2\,(g^{-1}X_Lg-X_R)h_2^{-1}+h_2X_Rh_2^{-1}]\,h_2>
\label{particulier}
\end{eqnarray}

The equation~(\ref{particulier}) decomposes into two independent equations
for the left and right translations, which written on $\cal S$ read:
\begin{eqnarray}
<L,Q^{-1}[X_L,V_L]Q>&=&<L,[Z,\nabla_g h_2\,(Q^{-1}X_LQ)]>\label{part.L}\\
<L,[X_R,V_R]>&=&<L,[Z,\nabla_g h_2\,(X_R)-X_R]>\label{part.R}
\end{eqnarray}

In order to further study these equations we first compute
$\nabla_g h_2\,(X)$ and $\nabla_g h_2\,(Q^{-1}XQ),$ $X\in \cal H$.
\begin{lem} \label{nablag}
We have on $\cal S$:
\begin{itemize}
\item{a)} $\nabla_g h_2\,(X)=X,\;\forall X \in \cal H$.
\item{b)} $\nabla_g h_2\,(Q^{-1}XQ)=D_Q(X),\;\forall X \in \cal H$.
\end{itemize}
where $D_Q(X)$ takes its values in $\cal B$ and vanishes on ${\cal H}_\mu$.
Moreover $D_Q(X)=X$ for $X \in \cal B$.
\end{lem}

\noindent{\it Proof.}
Right translations of an element of $N_\mu$ by
elements of $H$ always give elements on $N_\mu$ on which the decomposition
is unique. Hence $h_2(Q \exp (tX) )=h_2(Q).\exp (tX)$ directly
leading to $\nabla_g h_2\,(X)=X$ on $\cal S$.
Moreover due to Proposition~(\ref{unicite})
we have $h_2(hg)=d_g(h)h_2(g)$ with $d_g(h)\in B$ and
$d_g(h)=1$ if $(g,\xi) \in N_\mu$ and $h \in {\cal H}_\mu$. Taking $h$
infinitesimal yields the result. Finally if $X \in \cal B$ we have
$D_Q(X)=\nabla_g h_2\,(X)=X$ since $Q$ and $X$ commute.

Let us notice that equation~(\ref{part.R}) is identically satisfied for
any $V_R$ as it should be since the isotropy group is $H_\mu\times H$.
As a matter of fact $[X_R,V_R] \in \cal H$ while $L \in {\cal K}^*$ and
$\nabla_g h_2\,(X_R)-X_R=0$ by Lemma~(\ref{nablag}).

On the other hand  equation~(\ref{part.L}) reads:
\begin{equation}
<\mu\, ,\, [X,V_Z]>\, =\, <L\, ,\, [\, Z,D_Q(X)\, ]> \label{left}
\end{equation}
This is exactly equation~(\ref{eqvv}) for the appropriate function $F_Z$
and its solution is given by:
\begin{equation}
V_Z=\hat{\mu}^{-1}(\lambda_Z) {\rm ~where~}
\lambda_Z:\; X\to <L\, ,\, [Z,D_Q(X)]>\label{solution}
\end{equation}
Here $V_Z$ is an element of $({\cal H}/{\cal H}_\mu)$
depending linearly on $Z$.

We are now in a position to prove the existence of an $R$-matrix for the
Calogero model.
\begin{theo}\label{Rgene}
There exists a linear mapping $R:\;{\cal K}\to {\cal H}$ such that:
\begin{equation}
\{ \, L(X)\, ,\, L(Y)\, \}_{\rm reduced}=L\, (\,[X,RY]+[RX,Y]\,)
\label{matricer}
\end{equation}
and $R$ is given by:
\begin{equation}
R\,(X)= \nabla_g h_2\,(X) + {1 \over 2} D_Q\,( V_X)
\label{rformel}
\end{equation}
Hence the Calogero model is integrable.
\end{theo}

\noindent{\it Proof.} One uses the equation~(\ref{bienreduit}) and first
compute the unreduced Poisson bracket $\{F_X,F_Y\}$. Using
equations~(\ref{poi1},\ref{poi2},\ref{poi3}) one gets:
\begin{eqnarray*} \lefteqn{
\{ \xi(h_2^{-1}Xh_2),\xi(h_2^{-1}Yh_2) \}=-\xi(h_2^{-1}[X,Y]h_2)}\\
&&+\xi \left(h_2^{-1}\left(\, [X,\nabla_g h_2(h_2^{-1}Yh_2) h_2^{-1}]
+ [\nabla_g h_2(h_2^{-1}Xh_2) h_2^{-1},Y] \, \right) h_2 \right)
\end{eqnarray*}
Taking the value of this expression on $\cal S$ yields:
\begin{equation}
\{F_X,F_Y\}_{\, |\,{\cal S}}=\,<L, [X,\nabla_g h_2\, (Y)]+
[\nabla_g h_2\, (X),Y]>
\end{equation}
Here we have taken into account the fact that $X,Y \in \cal K$ hence
$[X,Y]\in \cal H$ and therefore $L([X,Y])=0$ since $L\in {\cal K}^*$.
We now evaluate the second term. Replacing $(V_Y.p).F_X=\lambda_{F_X}(V_Y)$
in equation~(\ref{bienreduit}) by its expression~(\ref{solution}) gives:
$${1\over 2}\left( \lambda_{F_X}(V_Y)-\lambda_{F_Y}(V_X) \right)=
{1\over 2}<L,[X,D_Q\,(V_Y)]+ [D_Q\,(V_X),Y]>$$
Adding the two terms yields the result.

Of course, due to the general theory $L\, (\,[X,RY]+[RX,Y]\,)$ does not
depend on the choice of the extension of $h_2$ out of $N_\mu$ but
the $R$--matrix depends on it through the choice of the function $D_Q$.
In order to get a simple form it is convenient to fix the choice of this
function.
As a consequence of  Lemma~(\ref{nablag}) all the indetermination reduces to
the value of $D_Q$ on the subspace $\cal C$ introduced in eq.~(\ref{hbc}),
and this can be chosen arbitrarily since $X.p$ is not tangent to $N_\mu$
when $p \in \cal S$ and $X \in \cal C$. The most natural choice is:
\begin{equation}
D_Q\,(X)=0, \quad X \in \cal C \label{choixdq}
\end{equation}
This choice has the important consequence:
\begin{prop}\label{simplification}
For any $Z \in \cal K$ we have $V_Z \in \cal C$ hence $D_Q\,(V_Z)=0$.
The $R$--matrix is then simply given by:
\begin{equation}
RZ=\nabla_g h_2\,(Z) \label{rsimple}
\end{equation}
\end{prop}

\noindent{\it Proof.}
Since $V_Z$ is defined by the equation~(\ref{left}) we see that
$\mu\,([X,V_Z])=0 \quad \forall X \in \cal C$. But $\cal C$ is a maximal
isotropic
subspace hence $V_Z \in \cal C$.

To proceed we need to compute the variation of the function $h_2(g)$
induced by a variation of $g$. It is given in the:
\begin{prop}\label{nablah2}
On the section $\cal S$ with $Q=\exp(q) \in A$ we have:\hfil\break
For $X \in \cal K$ i.e. $X=X_0+\sum X_\alpha (e_\alpha-\sigma e_\alpha),
\; X_0 \in \cal A$
\begin{equation}
\nabla_g h_2  ( X)=-h_0(X)+\sum_\alpha X_\alpha \coth (\alpha(q))
(e_\alpha+\sigma e_\alpha) \label{dh2}
\end{equation}
Here $h_0(X)$ is a linear function from $\cal G$ to $\cal B$ which
is fixed by the condition:
\begin{equation}
X_L \equiv h_0(X)-\sum_{\alpha }\,{X_\alpha \over \sinh \alpha(q)}
\,(e_\alpha+\sigma e_\alpha) \in {\cal H}_\mu \oplus \cal C
\label{mmain}
\end{equation}
\end{prop}

\noindent{\it Proof.} Since according to Proposition~(\ref{unicite})
any group element can be written in the form $g=h_1(g) Q(g) h_2(g)$ any
tangent vector $v$ to $G$ at $Q$ can be written {\em uniquely} as:
\begin{equation}
v=X_L. Q + Q .T + Q. X_R {\rm ~ with ~} X_L \in {\cal H}_\mu \oplus {\cal C},
\; T \in {\cal A},\; X_R \in \cal H \label{main}
\end{equation}
Let us remark that $X_L= \nabla_g h_1 (Q^{-1} v)$,
$X_R= \nabla_g h_2 (Q^{-1} v)$. If $v=X.Q$ with $X \in \cal B$
we have $X_L=0,\,T=0,\,X_R=X$ as in Lemma~(\ref{nablag}).
On the other hand if $v=X.Q$ with $X \in \cal C$ we take
$X_L=X,\,T=0,X_R=0$. Comparing with Lemma~(\ref{nablag})
we see that this is equivalent to $D_Q(X)=0, \; \forall X\in \cal C$
and we are in the situation described in equation~(\ref{choixdq}).
If $v=X.Q$ with $X \in {\cal H}_\mu$ we have
$X_L=X,\,T=0,X_R=0$. If $v=Q.X$ with $X\in\cal H$ one has
$X_L=0,\,T=0,\,X_R=X$ and finally if $v=Q.T$ with $T\in\cal A$ we have
$X_L=0,\,X_R=0$. We have completely described $\nabla_g h_1$ and
$\nabla_g h_2$ at the point $Q$ and we have found that that the
choice~(\ref{choixdq}) is equivalent to $\nabla_g h_1 \in
{\cal H}_\mu \oplus \cal C$.

Let us now assume that $v=Q.X$ with $X \in\cal K$. Decomposing
$v$ as in equation~(\ref{main}) we get $X=Q^{-1}. X_L. Q
+ T + X_R$ and writing $X_L=h_0(X)+\sum_{\alpha }\,h_\alpha(X)\,(e_\alpha
+\sigma e_\alpha)$ we have:
$$Q^{-1}. X_L. Q=h_0+\sum_{\alpha }\,h_\alpha \cosh \alpha(q)
\,(e_\alpha+\sigma e_\alpha)-\sum_{\alpha }\,h_\alpha \sinh \alpha(q)
\,(e_\alpha-\sigma e_\alpha)$$
so that projecting on $\cal H$ and $\cal K$ yields:
\begin{eqnarray*}
X &=& -\sum_{\alpha }\,h_\alpha \sinh \alpha(q) \,(e_\alpha-\sigma e_\alpha)
+T\\
0 &=& h_0+\sum_{\alpha }\,h_\alpha \cosh \alpha(q)\,(e_\alpha+\sigma
e_\alpha)+X_R
\end{eqnarray*}
This system is uniquely solved by:
$$T=X_0 \quad h_\alpha(X)=-{X_\alpha \over \sinh \alpha(q)}
\quad X_R=-h_0(X)+\sum_\alpha X_\alpha \coth (\alpha(q))
(e_\alpha+\sigma e_\alpha)$$
Notice that $h_0$ is uniquely determined by the condition
$X_L \in {\cal H}_\mu \oplus \cal C$ knowing the $h_\alpha$,
since $X_L$ is uniquely determined in equation~(\ref{main}).
This fixes the $R$--matrix corresponding
to the choice~(\ref{choixdq}). We shall in the next section
compute $R$ in a concrete case by applying Proposition~(\ref{nablah2}).

\section{The $R$--matrix of the standard Calogero model}

\setcounter{equation}{0}

The standard Calogero model can be obtained as above starting from
the non compact group $G=SL(n,\Bbb{C})$ and its maximal compact
subgroup $H=SU(n)$ as first shown by~\cite{OlPe76}. We
choose the momentum $\mu_L$ as described in Section~(3.1) so that
the isotropy group $H_\mu$ be a maximal proper Lie subgroup of $H$.
Obviously one can take $\mu_L$ of the form:
\begin{equation}
\mu_L= i\,(vv^+ -1) \label{musln}
\end{equation}
where $v$ is a vector in $\Bbb{C}^n$ such that $v^+ v=1$, hence $\mu_L$
is a traceless antihermitian matrix. Then $g \mu_L g^{-1} =\mu_L$ if
and only if $g v= c v$ where $c$ is a complex number of modulus 1. Hence
$H_\mu = S(U(n-1)\times U(1))$ which has the above--stated property.

In this case the automorphism $\sigma$ is given by $\sigma\,(g)=
(g^+)^{-1}$ (notice that we consider only the real Lie group structure),
$B$ is the group of  diagonal matrices of determinant 1 with
pure phases on the diagonal and $A$ is the group of real diagonal
matrices with determinant 1. The property~(\ref{2.cap})
is then satisfied as soon as the vector $v$ has no zero component.
As a matter of fact, $v$ is further constrained by $\mu_L$
being a value of the
moment map. Considering equation~(\ref{2.mu}) we see that $\mu_L$ has
no diagonal element, which implies that all the components of $v$
are pure phases $v_j=\exp (i\theta_j)$.
These extra phases which will appear in the Lax matrix
can however be conjugated out by the adjoint action of a constant matrix
${\rm diag}\,(\exp (i\theta_j))$ hence we shall from now on set $v_j=1$
for all $j$. This is the solution first considered by Olshanetskii
and Perelomov.

We now show that the reduced phase space has the correct dimension
$2\,{\rm dim}\,{\cal A} = 2\,{\rm dim}\,\cal B$. Counting {\em real}
parameters any $g \in SL(n,\Bbb{C})$ involves $(2n^2-2)$ parameters
(notice that ${\rm det}\,g =1$ gives 2 conditions) so that $T^*G$
involves $2(2n^2-2)$ parameters.
The surface $N_\mu$ in $T^*G$ is defined by $2\,{\rm dim}\,\cal H$
equations namely $P_{{\cal H}^*} \xi =0$ and $P_{{\cal H}^*}
\,{\rm Ad}^*_g \xi=\mu_L$. Since ${\rm dim}\,{\cal H}=n^2-1$ we see that
$N_\mu$ is of dimension $(2n^2-2)$. Finally $H_\mu$ is of dimension
$(n-1)^2-1+1$ and $H_\mu \times H$ of dimension $(2n^2-2n)$, hence
the reduced phase space is of dimension $(2n-2)$ which exactly
corresponds to the Calogero model.
\begin{prop}
\label{decomposition}
We have:
$${\cal H}_\mu=\{M \vert\; M^+=-M ,\; {\rm Tr}\,M=0, \;Mv=0 \}\oplus
i\Bbb{R}\, {\bf \mu_L } $$
One can take ${\cal C}=({\cal H}_\mu \oplus {\cal B})^\perp$ (here we take
the orthogonal under the Killing form) so that:
$${\cal C}=\{M \vert\; M^+=-M ,\; M_{ij}=u_i-u_j, \; u_i \in \Bbb{R}, \;
\sum u_i=0 \}$$
Finally  $\cal B$ and $\cal C$ are  a pair of maximal isotropic
subspaces of ${\cal H}/{\cal H}_\mu$.
\end{prop}

\noindent{\it Proof.}
First of all $Mv=0$ is
equivalent to $v^T M=0$ since $v$ is real and the Killing form is simply
$(X,Y)={\rm Tr}(XY)$. Hence the orthogonal of the space of matrices
such that $Mv=0$ is the space $\{ (u v^T - v u^T)|\; \forall u \in \Bbb{C}^n
\}$.
We then ask that such a matrix be orthogonal to any element of ${\cal
B}$.
This immediately gives $u_i-u_i^* =\lambda \; \forall i$. Finally we
ask that this matrix be orthogonal to $\mu_L$. This implies $\lambda
=0$. Hence this matrix
takes the form $M_{ij}=u_i-u_j \; u_i \in {\Bbb R}$ and one can set $\sum
u_i=0$.

The skew--symmetric form $\chi$ on ${\cal B}\oplus {\cal C}$
can be written:
$$\chi\,(X,Y)=\,<\mu_L , [X,Y] > \,=i v^+[X,Y]v$$
leading to:
$$ \chi\,(X,Y)= 2 i n \sum_i \,(\rho_i \gamma_i -\kappa_i \beta_i)$$
where $X=\rho + \beta$, $Y=\kappa + \gamma$, $\rho_{ij}=\rho_i \delta_{ij}$,
$\beta_{ij}=\beta_i -\beta_j$, $\kappa_{ij}=\kappa_i \delta_{ij}$,
$\gamma_{ij}=\gamma_i -\gamma_j$ and $\sum \rho_i=\sum \beta_i=
\sum \kappa_i = \sum_i \gamma_i=0$.
Hence $\cal B$ and $\cal C$ are a pair of complementary Lagrangian subspaces.

The root vectors appearing in the radicial decomposition~(\ref{radix})
are the $E_{kl}$ and $iE_{kl}$ with $k \neq l$ where $(E_{kl})_{ab}=
\delta_{ka}\delta_{lb}$ and $\sigma(E_{kl})=-E_{lk},\;
\sigma(iE_{kl})=iE_{lk}$. In this basis $\mu_L$  given by~(\ref{musln})
reads $\mu_L= \sum_{k < l}\, (i E_{kl} + i E_{lk} )$.

The Lax matrix $L$ is then given by Proposition~(\ref{laxpair})
and therefore
$$L=p+\sum_{k<l}\, {1\over \sinh(q_k-q_l)} (i E_{kl} - i E_{lk})$$
More explicitely:
$$L= \pmatrix{
p_1 & & \cr
&  & {i \over \sinh(q_i -q_j)}  \cr
& \ddots & \cr
{i \over \sinh(q_j -q_i)} & & \cr
& & p_n \cr}$$

The $R$--matrix can now be deduced straightforwardly from
Proposition~(\ref{nablah2}). We compute $RX$ for $X\in \cal K$
of the form:
$$X=\sum_{k<l}\,  x_{kl} (E_{kl}+E_{lk}) + \sum_{k<l}\, y_{kl}
(iE_{kl}-iE_{lk})
+\sum_k \, z_k E_{kk},\quad x_{kl} ,\,y_{kl},\, z_k\in\Bbb{R}$$
The element $X_L$ appearing in equation~(\ref{mmain}) reads:
$$X_L=h_0(X)+\sum_{k<l}\,{1 \over \sinh (q_k-q_l)}\,\left[\vphantom{1\over 2}
-(x_{kl}+iy_{kl})E_{kl}+(x_{kl}-iy_{kl})E_{lk} \right]$$
where $h_0$ is a pure imaginary diagonal traceless matrix.
The matrix $X_L$ belongs to ${\cal H}_\mu \oplus \cal C$ and this condition
uniquely determines $h_0$. Since the action of an element $M_{ij}=
u_i-u_j$ of $\cal C$ on the vector $v$ definining $\mu_L$ is $Mv=nu$
and the action of an element of ${\cal H}_\mu$ on $v$ gives $i \theta v,\;
\theta \in \Bbb{R}$ we get the condition $\sum_l \,(X_L)_{kl}=i \theta
+ nu_k$. Separating the real and imaginary parts in this equation the real
part immediately determines the $u_i$ which are of no concern to us, and
the imaginary part gives:
$$h_0(X)_{kk}=i \theta +i \sum _{l>k} {{y_{kl}} \over \sinh (q_k-q_l)}
-i \sum_{l<k} {y_{lk} \over \sinh (q_k-q_l)}$$
Of course $\theta$ is determined by $\sum_k \, (h_0)_{kk}=0$. Finally
equation~(\ref{dh2}) produces the $R$--matrix:
\begin{equation}
RX=-h_0(X)+\sum_{k<l}\,{\coth (q_k-q_l)}\,\left[\vphantom{1 \over 2}
(x_{kl}+iy_{kl})E_{kl}-(x_{kl}-iy_{kl})E_{lk} \right]
\label{rsln}
\end{equation}

In order to recognize the form of the $R$--matrix first found in~\cite{AvTa93}
we write $RX={\rm Tr}_2 \, R_{12}.{\bf 1}\otimes X$ and we find:
$$R_{12}=\sum_{k\neq l}\, \coth (q_k-q_l) E_{kl}\otimes E_{lk}
+ {1 \over 2} \sum_{k \neq l}\, {1 \over \sinh (q_k-q_l)}
(E_{kk}-{1 \over n}\,{\bf 1})\otimes (E_{kl}-E_{lk})$$
This is exactly the correct $R$--matrix of the Calogero model for the
potential $1/\sinh(x)$, and the other potentials $1/\sin(x)$ and
$1/x$ have similar $R$--matrices obtained by analytic continuation.

\section{The $R$--matrix of the $SU(n,n)$ Calogero model.}

\setcounter{equation}{0}

The $SU(n,n)$ Calogero model is obtained by starting from the non compact group
$G=
SU(n,n)$. This is the subgroup of $SL(2n,{\Bbb C})$ which leaves invariant the
sesquilinear quadratic form defined by
\begin{eqnarray}
Q((u_1,v_1),(u_2,v_2)) = \pmatrix{u_1^+ & v_1^+} J
\pmatrix{u_2 \cr v_2}= u_1^+ v_2 + v_1^+ u_2
\label{quadra}
\end{eqnarray}
where $u_i,v_i$ are vectors in ${\Bbb C}^n$ and $J$ is the matrix
\begin{eqnarray}
J= \pmatrix{0 & {\bf 1} \cr {\bf 1} & 0}.
\nonumber
\end{eqnarray}

The Lie algebra of $SU(n,n)$ therefore
consists of block matrices
\begin{eqnarray}
{\cal G}= \{ \pmatrix{a & b \cr c & d } \vert a=-d^+ ,\; {\rm Tr}\; (a+d)=0,
\; b^+ = -b,\; c^+ = -c \}
\label{sunn}
\end{eqnarray}
where $a,b,c,d$ are $n \times n$ complex matrices.

We consider again the automorphism $\sigma: \sigma(g)=(g^+)^{-1}$, which can be
consistently restricted to $SU(n,n)$. Its fixed points at the Lie algebra level
consist of block matrices
\begin{eqnarray}
{\cal H}=\{ \pmatrix{a & c \cr c & a} \vert a^+ =-a, \; { \rm Tr } \,  (a)=0,\;
c^+ =-c \}
\label{sunun}
\end{eqnarray}
This Lie algebra is isomorphic to the Lie algebra of $S(U(n) \times U(n))$, the
two
$u(n)$'s being realized respectively by $a+c$ and $a-c$.

The subalgebra ${\cal B}$ consists of matrices of the form (\ref{sunun}) with
$c=0$, and $a$ is a diagonal matrix of zero trace and purely imaginary
coefficients. The Abelian subalgebra ${\cal A}$ consists of matrices of the
form
(\ref{sunn}) with $b=c=0$ and $a=-d$ is a real diagonal matrix.

To perform the reduction, we choose as above $\mu^R=0$ and
\begin{eqnarray}
\mu^L = i(v v^+ - {\bf 1})+ i\gamma J
\label{momentsu}
\end{eqnarray}
The vector $v$ has $2n$ components all equal to 1, and $J$ is the matrix
defining
the quadratic form (\ref{quadra}). Remark that $J$ is invariant under the
adjoint
action of ${\cal H}$. Then $g\mu^L g^{-1} =\mu^L\;\forall  g \in H$ is
equivalent to $gv =e^{i \theta }v$. Writing an element of ${\cal H}$
as
\begin{eqnarray}
\pmatrix{ u+w & u-w +i \lambda {\bf 1} \cr u-w +i \lambda {\bf 1} & u+w };\;
u^+ =-u, \; w^+=-w;\;
{ \rm Tr } \, (u)={ \rm Tr } \, (w)=0.
\nonumber
\end{eqnarray}
the subalgebra ${\cal H}_\mu$ consists of matrices
\begin{eqnarray}
\pmatrix{ \tilde{u}+w & \tilde{u}-w +i \lambda {\bf 1}\cr \tilde{u}-w +i
\lambda {\bf 1} & \tilde{u}+w };\; \tilde{u}^+ =-u, \; w^+=-w;\;
{ \rm Tr } \, (\tilde{u})={ \rm Tr } \, (w)=0.
\label{HH}
\end{eqnarray}
and $\tilde{u}\tilde{v} = i \theta \tilde{v}$ where $\tilde{v}$ is an
$n$ component vector with all entries equal to one. This brings
us back to the $SL(n,\Bbb C)$ case.

Hence $H_\mu
=SU(n-1)\times U(1) \times SU(n) \times U(1)$.
The factor $SU(n)\times U(1)$ is generated by $w+ i \lambda {\bf
1}$, while the
factor generated by $SU(n-1)\times U(1)$ is generated by
$\tilde{u}$.
The isotropy group $H_\mu$ is indeed a maximal subgroup of
$S(U(n)\times U(n))$.

Notice that in equation (\ref{momentsu})
the parameter $\gamma$ is an arbitrary real number. This will
lead to existence of a second coupling constant in the
corresponding Calogero model.

We now compute the dimension of the reduced phase space. The
real dimension of $SU(n,n)$ is $4n^2 -1$, and so ${\rm
dim}\,T^*G = 8n^2 -2$. The dimension of $H= S(U(n) \times U(n)$
is $2n^2 -1$. Hence, the dimension of ${\cal N}_\mu$ is
$4n^2$. Now ${\rm dim}\,G_\mu = 4n^2 -2n$ and therefore the
dimension of the phase space is $2n=2\, {\rm dim}\,{\cal A}$ as it
should.

\begin{prop}
We have the decomposition
\begin{eqnarray}
{\cal H}= {\cal H}_\mu \oplus {\cal B} \oplus {\cal C}
\nonumber
\end{eqnarray}
where one can take ${\cal C} = ({\cal H}_\mu \oplus {\cal B})^\perp$
so that
\begin{eqnarray}
{\cal C}=\{ M \vert M =\pmatrix{ c & c \cr c & c};\; c_{ij}=u_i
-u_j,\; u_i \in {\Bbb R},\; \sum u_i =0 . \}
\nonumber
\end{eqnarray}
Moreover ${\cal B}$ and ${\cal C}$ are a pair of maximally isotropic
subspaces of the bilinear form $\chi$ defined in (\ref{chimu}).
\end{prop}
\noindent {\it Proof.} In the parametrization (\ref{HH}) of
${\cal H}$ we decompose the matrix $u$ according to proposition
(\ref{decomposition})
\begin{eqnarray}
u=\tilde{u} + iD + c
\nonumber
\end{eqnarray}
where $\tilde{u}\tilde{v}=i\theta \tilde{v}$, $D$ is a real
traceless diagonal matrix and $c_{ij}=u_i-u_j$ with $u_i$ real.
Hence, any element
of ${\cal H}$ can be written as
\begin{eqnarray}
\pmatrix{ \tilde{u}+w' & \tilde{u}-w' +i \lambda {\bf 1} \cr
\tilde{u}-w' +i \lambda {\bf 1} & \tilde{u}+w' }
+ \pmatrix{2iD & 0 \cr 0 & 2iD}
+ \pmatrix{c & c \cr c & c}
\nonumber
\end{eqnarray}
where $w'=w-iD$. The first matrix parametrizes ${\cal H}_\mu$ as
in equation (\ref{HH}). The second matrix parametrizes ${\cal
B}$ and the third matrix parametrizes a supplementary subspace
${\cal C}$ of dimension $n-1$. The rest of the proof is
identical to the $SL(n,\Bbb C)$ case.

The root vectors in~(\ref{radix}) are $(i,j = 1\cdots n)$:
\begin{eqnarray}
E_{ij}-E_{j+n,i+n} && iE_{ij}+iE_{j+n,i+n}~~i\neq j \nonumber \\
E_{i,j+n}-E_{j,i+n}&& iE_{i,j+n}+iE_{j,i+n} \nonumber \\
E_{i+n,j}-E_{j+n,i}&& iE_{i+n,j}+iE_{j+n,i} \nonumber
\end{eqnarray}
The automorphism $\sigma$ is the same as before: $\sigma(E_{kl})=-E_{lk}$ and
$\sigma(iE_{kl})= iE_{lk}$. In this basis the momentum
$\mu_L$ reads
\begin{eqnarray}
\mu_L &=&\sum_{i<j}(1+\sigma)(iE_{ij}+iE_{j+n,i+n}) \nonumber \\
&&+ \sum_{i<j}(1+\sigma)(iE_{i,j+n}+iE_{j,i+n}) \nonumber \\
&&+(\gamma +1)\sum_i (1+\sigma)(iE_{i,i+n}) \nonumber
\end{eqnarray}
Then, from proposition (\ref{laxpair}), the Lax matrix becomes
\begin{eqnarray}
L&=& p +\sum_{i<j}{1\over \sinh (q_i -q_j)}(1-\sigma)
(iE_{ij}+iE_{j+n,i+n}) \nonumber \\
&&+  \sum_{i<j}{1\over \sinh (q_i +q_j)}(1-\sigma)
(iE_{i,j+n}+iE_{j,i+n}) \nonumber \\
&&+ (\gamma +1)\sum_i{1\over \sinh (2 q_i )} (1-\sigma)(iE_{i,i+n})
\nonumber
\end{eqnarray}
where $p$ is a generic element of ${\cal A}$ of the form ${\rm
diag}\, p_i,-{\rm diag}\, p_i$.

The $R$-matrix is then computed straightforwardly:
\begin{eqnarray}
R_{12}&=&{1\over 2} \sum_{k\neq l} \coth (q_k-q_l) (E_{kl}+E_{k+n,l+n})
\otimes (E_{lk}-E_{l+n,k+n}) \nonumber \\
&+&{1\over 2} \sum_{k,l} \coth (q_k +q_l) (E_{k,l+n}+E_{k+n,l})
\otimes (E_{l+n,k}-E_{l,k+n}) \nonumber \\
&+& {1\over 2} \sum_{k\neq l}{1\over \sinh (q_k -q_l)}
(E_{kk}+E_{k+n,k+n}-{1\over n} {\bf 1})\otimes (E_{kl}-E_{k+n,l+n})
\nonumber \\
&+&{1\over 2} \sum_{k,l}{1\over \sinh (q_k+q_l)}
(E_{kk}+E_{k+n,k+n}-{1\over n} {\bf 1})\otimes (E_{k,l+n}-E_{k+n,l})
\nonumber
\end{eqnarray}

Using this result, one can compute the $M$ operator in the Lax
equation from formula (\ref{RdH}). We get
\begin{eqnarray}
M&=&\sum_{k\neq l} {\cosh(q_k-q_l)\over \sinh^2 (q_k-q_l)}
(E_{kl}+E_{k+n,l+n})+\sum_{k,l} {\cosh(q_k+q_l)\over \sinh^2
(q_k+q_l)}(E_{k,l+n}+E_{k+n,l}) \nonumber \\
&+& \gamma\sum_{k} {\cosh(2 q_k)\over \sinh^2 (2 q_k)}(E_{k,k+n}
+E_{k+n,k})\nonumber \\
&+& \sum_{k\neq l}\left({1\over
\sinh^2(q_k-q_l)}+{1\over \sinh^2 (q_k+q_l)}\right)(E_{kk}+E_{k+n,k+n}-{1\over
n}{\bf 1}) \nonumber \\
&+& \gamma\sum_{k} {1\over \sinh^2(2q_k)} (E_{kk}+E_{k+n,k+n}-{1\over n}
{ \bf 1})
\nonumber
\end{eqnarray}
This is precisely the $M$ matrix found by Olshanetsky and Perelomov
\cite{OlPe81}.

\end{document}